\documentclass[12pt]{elsart}

\usepackage{graphicx}
\usepackage{subfigure}
\usepackage{amsmath}
\usepackage{cite}

\journal{Physics Letters A}

\begin{document}

\begin{frontmatter}

\title{Critical behavior of a triangular lattice Ising AF/FM bilayer}
\author{M. \v{Z}ukovi\v{c}\corauthref{cor}},
\ead{milan.zukovic@upjs.sk}
\author{A. Bob\'ak}
\address{Department of Theoretical Physics and Astrophysics, Faculty of Science,\\ 
P. J. \v{S}af\'arik University, Park Angelinum 9, 041 54 Ko\v{s}ice, Slovak Republic}
\corauth[cor]{Corresponding author.}

\begin{abstract}
We study a bilayer Ising spin system consisting of antiferromagnetic (AF) and ferromagnetic (FM) triangular planes, coupled by ferromagnetic exchange interaction, by standard Monte Carlo and parallel tempering methods. The AF/FM bilayer is found to display the critical behavior completely different from both the single FM and AF constituents as well as the FM/FM and AF/AF bilayers. Namely, by finite-size scaling (FSS) analysis we identify at the same temperature a standard Ising transition from the paramagnetic to FM state in the FM plane that induces a ferrimagnetic state with a finite net magnetic moment in the AF plane. At lower temperatures there is another phase transition, that takes place only in the AF plane, to different ferrimagnetic state with spins on two sublattices pointing parallel and on one sublattice antiparallel to the spins on the FM plane. FSS indicates that the corresponding critical exponents are close to the two-dimensional three-state ferromagnetic Potts model values.
\end{abstract}


\end{frontmatter}

\section{Introduction}
Magnetic bilayers and multilayers are of considerable theoretical interest since they allow studying the cross-over phenomena between the two- and three-dimensional systems~\cite{cape76,jong90}. On the experimental side, recent techniques facilitate fabricating a variety of such layered structures in a highly controlled and tunable way~\cite{jong76,bala08,shi15}, which can lead to useful technological applications such as magneto-optical discs~\cite{shim92}.\\
\hspace*{5mm} Properties of Ising bilayers consisting of two different ferromagnetic layers coupled by either ferromagnetic (FM) or antiferromagnetic (AF) exchange interactions of varying strengths have been investigated in a number of studies~\cite{ferr91,hans93,hori97,lipo98,li01,kim01,ghae04,monr04,szal12,szal13}. Regarding their critical behavior, it has been found that they belong to the same universality class as a two-dimensional Ising model and their critical temperature is controlled by the so call shift exponents that depends on the interlayer to intralayer interaction ratio. In the absence of an external magnetic field due to symmetry reasons these findings apply to both FM and AF layered systems, as long as the lattice is bipartite.\\
\hspace*{5mm} Apparently, the situation is completely different if one considers an AF bilayer on a nonbipartite, such as triangular, lattice. A single layer, i.e. a two-dimensional triangular lattice Ising antiferromagnet (TLIA) is exactly known to show no long-range ordered (LRO) phase down to zero temperature due to high geometrical frustration~\cite{wann50}. A recent study of cross-over phenomena in a layered system obtained by stacking of individual TLIA layers on top of each other revealed a rather exotic behavior termed ``stiffness from disorder'' leading to a low temperature reentrance of two Berezinskii-Kosterlitz-Thouless transitions~\cite{lin14}. Nevertheless, this phenomenon is only observed in multilayer systems exceeding a certain critical number of layers but not in the bilayer. The latter shows critical properties  similar to the two-dimensional TLIA.\\
\hspace*{5mm} Considering the above, we find it interesting to study an Ising bilayer system consisting of one AF and one FM triangular planes. As mentioned above, the critical behavior of the decoupled planes is very distinct. While the AF one shows no LRO down to zero temperature due to high geometrical frustration, the FM one displays a single standard Ising universality class phase transition to the FM LRO phase. In the present Letter we show that in the system consisting of the coupled AF/FM planes the competing ordering and disordering tendencies enforced by the respective layers result in the critical behavior completely different from both the separate FM and AF planes as well as the FM/FM and AF/AF bilayers. 

\section{Model and Methods}
The model Hamiltonian can be written as
\begin{equation}
\label{Hamiltonian}
H=J_{\rm A}\sum_{\langle i \in {\rm A},j \in {\rm A} \rangle}\sigma_{i}\sigma_{j}-J_{\rm B}\sum_{\langle k \in {\rm B},l \in {\rm B} \rangle}\sigma_{k}\sigma_{l}-J_{\rm AB}\sum_{\langle i \in {\rm A},k \in {\rm B} \rangle}\sigma_{i}\sigma_{k},
\end{equation}
\noindent
where $\sigma_{i}=\pm 1$ is an Ising spin on the $i$th lattice site, the first two sums run over nearest neighbors (NN) within A and B planes, coupled by the exchange interactions $J_{\rm A}$ and $J_{\rm B}$, respectively, and the third sum runs over NN between the planes A and B, coupled by the exchange interaction $J_{\rm AB}$. In the following we will restrict our self to the fully isotropic case of $J_{\rm A}=J_{\rm B}=J_{\rm AB} \equiv J>0$.\\
\hspace*{5mm} In order to obtain temperature dependencies of various quantities we use standard Monte Carlo (MC) simulations following the Metropolis dynamics. Linear lattice sizes of the individual planes range from $L=24$ up to $168$ and periodic (open) boundary conditions are applied within (out of) planes. Simulations start from high temperatures and random initial states. Then the temperature is gradually lowered with the step $k_B\Delta T/J=0.05$ and simulations at the next temperature are initiated using the final configuration obtained at the previous temperature. For thermal averaging we use $10^5$ MC sweeps (MCS), after discarding another $2 \times 10^4$ MC for thermalization. Error estimates are obtained from three independent simulation runs. \\
\hspace*{5mm} In the low-temperature region, which is potentially problematic for frustrated spin systems due to time-consuming tunneling through multimodal energy landscapes resulting in extremely slow relaxation, we verify the reliability of the obtained results by applying the parallel tempering (PT) or replica exchange method~\cite{huku96}. The method overcomes energy barriers by a random walk in temperature space and allows exploration of complex energy landscapes of frustrated systems. We roughly tune the simulation temperature set by preliminary runs monitoring replica-exchange acceptance rates. For each lattice size, replica swaps at neighboring temperatures are proposed after each of $10^6$ MCS. \\
\hspace*{5mm} To obtain critical exponent ratios, we perform finite-size scaling (FSS) analysis, in which case we apply the reweighing techniques~\cite{ferr88}. The reweighing is performed at lattice-size-dependent pseudo-critical temperatures, estimated from the standard MC simulations, using $10^7$ MCS for statistical averaging and errors are estimated by applying the more reliable and precise $\Gamma$-method~\cite{wolf04}. \\
\hspace*{5mm} We measure the following basic thermodynamic quantities: The internal energy per spin
\begin{equation}
\label{ene}
e = \langle H \rangle/L^2,
\end{equation}
the magnetizations per spin of the separate planes A and B
\begin{equation}
\label{mag}
(m_{\mathrm{A}},m_{\mathrm{B}}) = (\langle M_{\mathrm{A}}\rangle,\langle M_{\mathrm{B}}\rangle)/L^2 = \Big(\Big\langle\Big|\sum_{i \in A}\sigma_{i}\Big|\Big\rangle, \Big\langle\Big|\sum_{j \in B}\sigma_{j} \Big|\Big\rangle\Big)/L^2,
\end{equation}
the staggered magnetization per spin (order parameter) within the AF plane A
\begin{equation}
\label{mag_s}
m_{s} = \langle M_{s} \rangle/L^2 = 3\Big\langle \max_{i=1,2,3}(M_{\mathrm{A},i})-\min_{i=1,2,3}(M_{\mathrm{A},i}) \Big\rangle/2L^2,
\end{equation}
where $M_{\mathrm{A},i}$ ($i=1,2,3$) are the three sublattices within the AF plane A and $\langle\cdots\rangle$ denotes thermal average. 
From the above quantities we further calculate the specific heat per site
\begin{equation}
\label{eq.c}c=\frac{\langle H^{2} \rangle - \langle H \rangle^{2}}{L^2k_{B}T^{2}},
\end{equation}
the susceptibility per site $\chi_{x}$, corresponding to the parameter $M_x$, $x=\mathrm{A},\mathrm{B},s$.
\begin{equation}
\label{eq.chi}\chi_{x} = \frac{\langle M_{x}^{2} \rangle - \langle M_{x} \rangle^{2}}{L^2k_{B}T}, 
\end{equation}
and the derivative and logarithmic derivative of $\langle M_x \rangle$ with respect to $\beta=1/k_{B}T$
\begin{equation}
\label{eq.D0}dm_x = \frac{\partial}{\partial \beta}\langle M_x \rangle = \langle M_x H \rangle- \langle M_x \rangle \langle H \rangle,
\end{equation}
\begin{equation}
\label{eq.D1}dlm_x = \frac{\partial}{\partial \beta}\ln\langle M_x \rangle = \frac{\langle M_x H \rangle}{\langle M_x \rangle}- \langle H \rangle.
\end{equation}
In the FSS analysis we employ the following scaling relations:
\begin{equation}
\label{eq.scal_chi}\chi_{x,max}(L) \propto L^{\gamma_x/\nu_x},
\end{equation}
\begin{equation}
\label{eq.scal_dlm}dlm_{x,max}(L) \propto L^{1/\nu_x},
\end{equation}
\begin{equation}
\label{eq.scal_dm}dm_{x,max}(L) \propto L^{(1-\beta_x)/\nu_x},
\end{equation}
\begin{equation}
\label{eq.scal_c}c_{max}(L) \propto L^{\alpha_x/\nu_x}, (\alpha \neq 0)
\end{equation}
\begin{equation}
\label{eq.scal_c1}c_{max}(L) = c_0+c_1\ln(L), (\alpha=0)
\end{equation}
\begin{equation}
\label{eq.scal_Tc}\beta_{x,max}(L) = \beta_c + a_x L^{-1/\nu_x},
\end{equation}
where $\beta_c$ is the inverse transition temperature and $\beta_{x,max}(L)$ are the inverse pseudo-transition temperatures, estimated as positions of the maxima of the above functions for a given $L$.

\section{Results and discussion}
\begin{figure}[t]
\centering
	\subfigure{\includegraphics[scale=0.5]{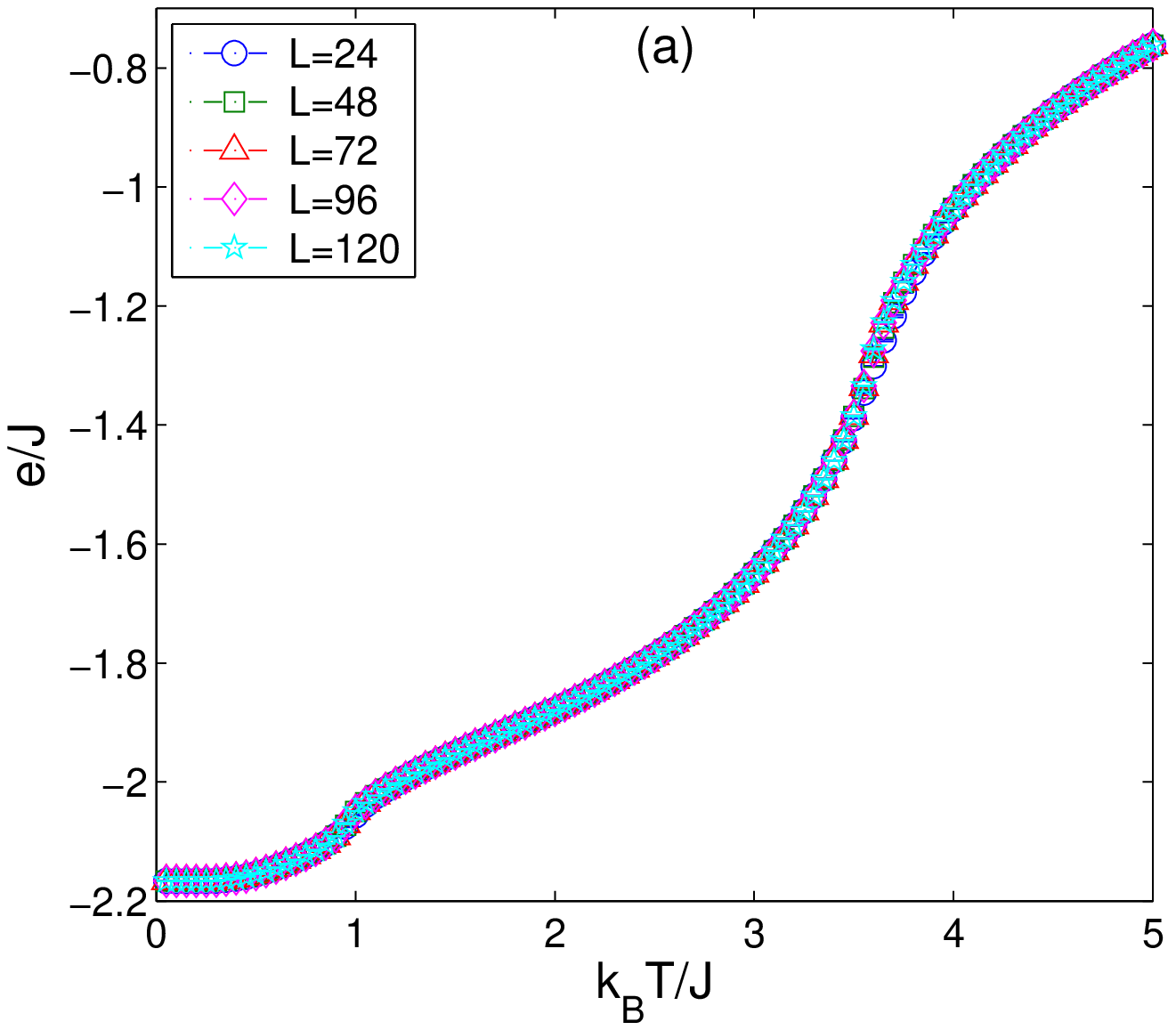}\label{fig:e-T}}
	\subfigure{\includegraphics[scale=0.5]{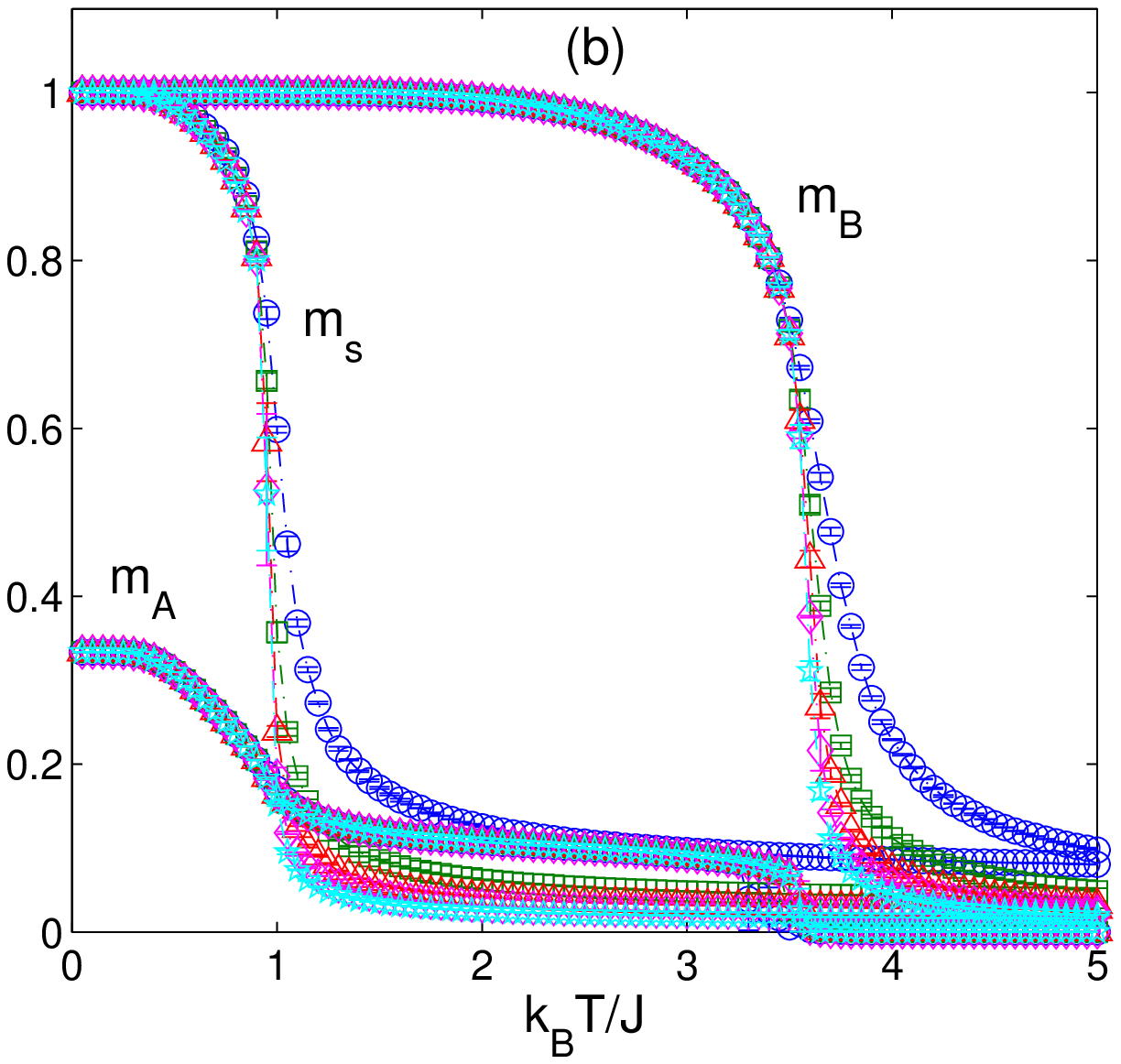}\label{fig:m-T}}
	\subfigure{\includegraphics[scale=0.5]{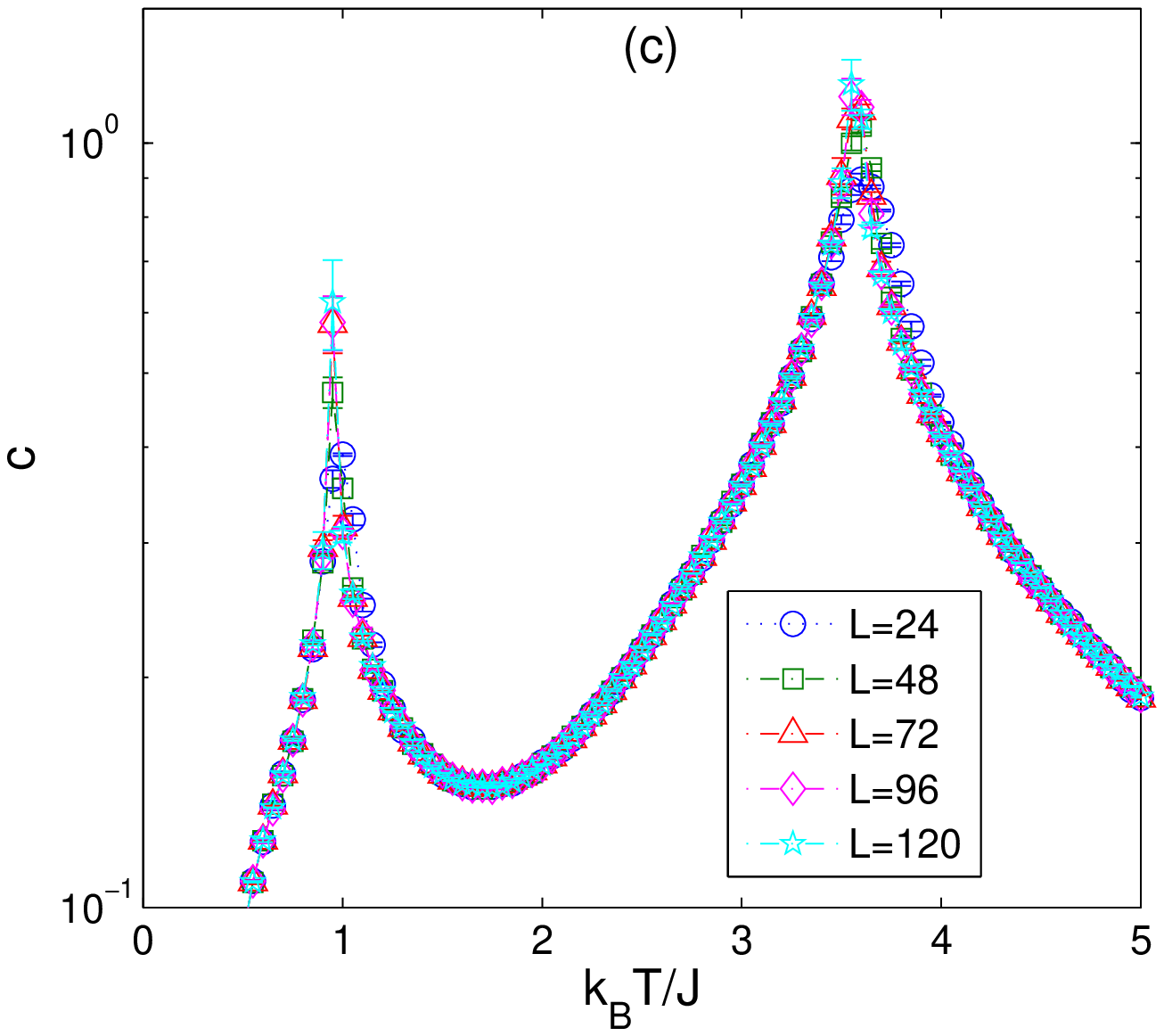}\label{fig:c-T_L24-120_log}}
	\caption{Temperature variations of (a) the internal energy $e/J$, (b) the magnetizations $m_{\mathrm{A}}$, $m_{\mathrm{B}}$ and $m_s$, and (c) the specific heat $c$, plotted for $L=24-120$.}\label{fig:T-x1}
	\end{figure}

\begin{figure}[t]
\centering
	\subfigure{\includegraphics[scale=0.5]{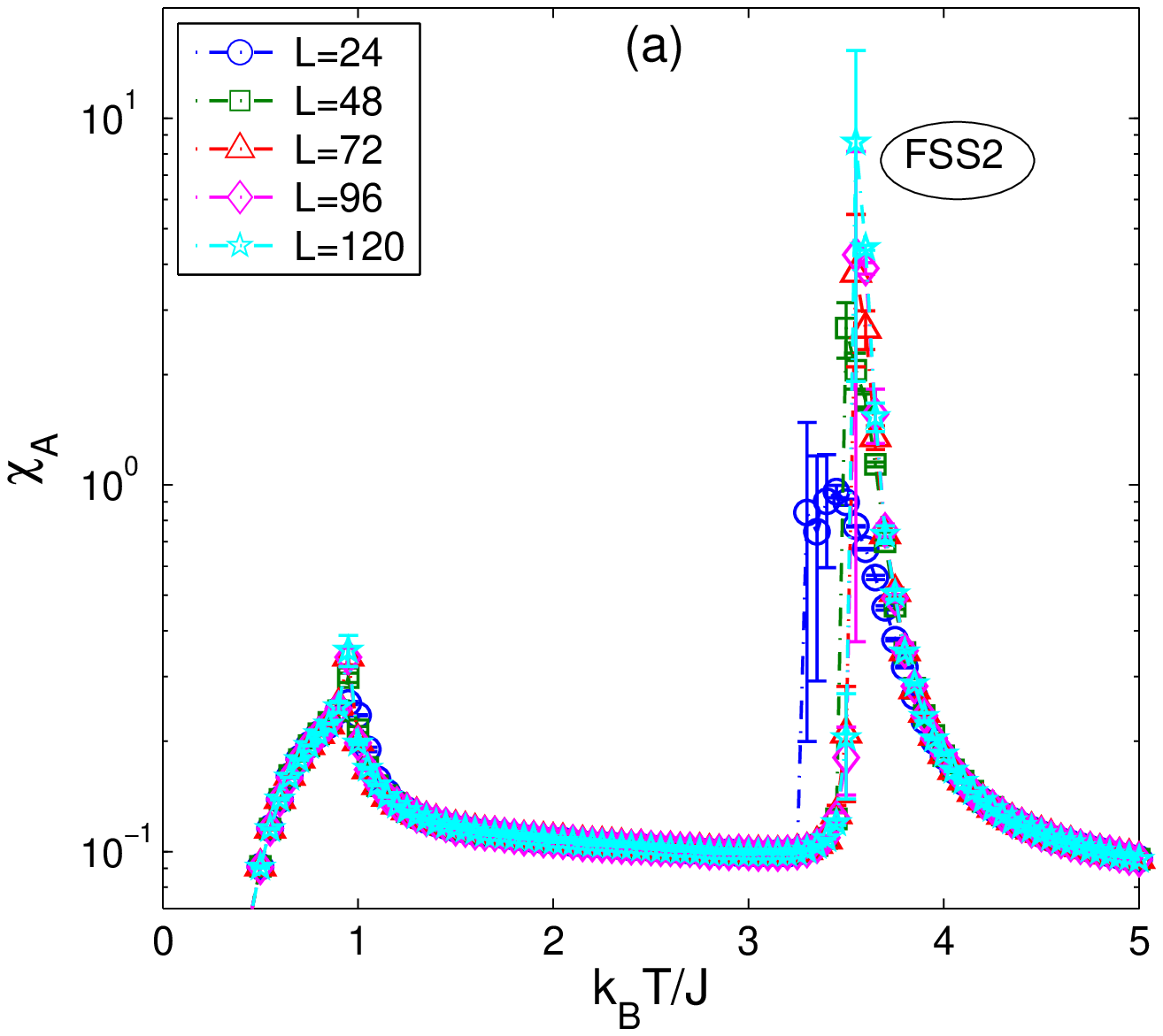}\label{fig:chiA-T_L24-120_log}}
	\subfigure{\includegraphics[scale=0.5]{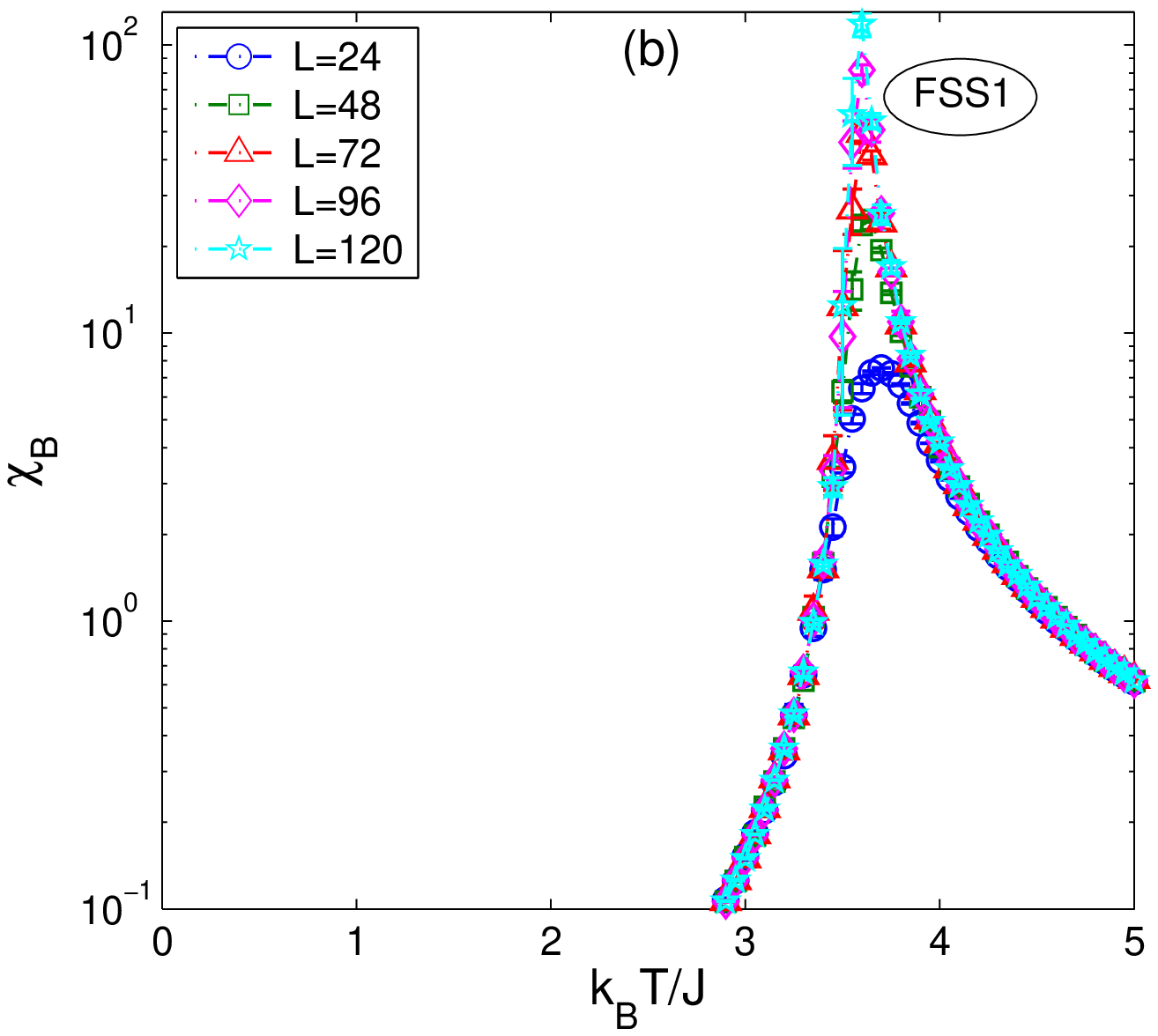}\label{fig:chiB-T_L24-120_log}}
	\subfigure{\includegraphics[scale=0.5]{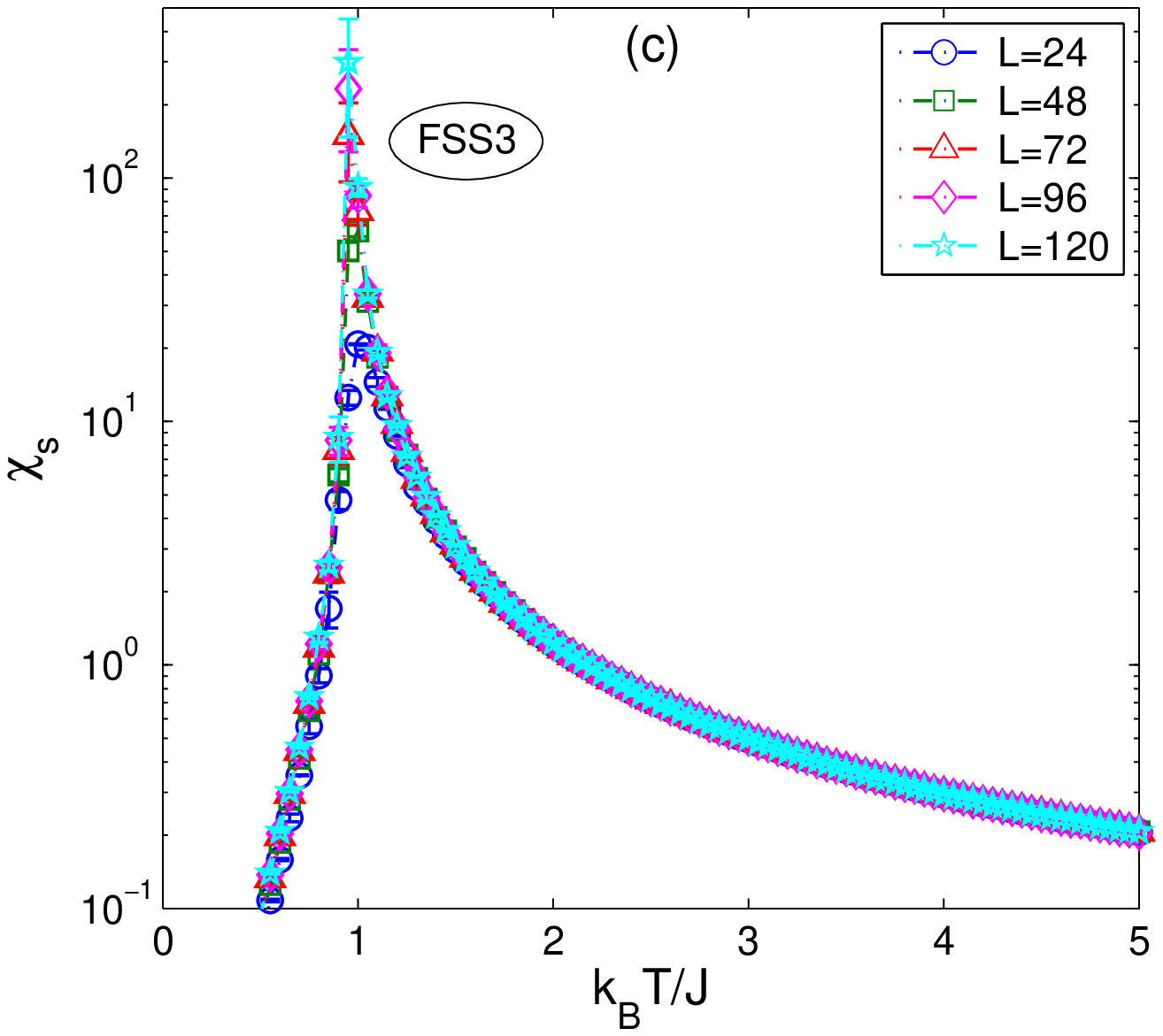}\label{fig:chis-T_L24-120_log}}
	\subfigure{\includegraphics[scale=0.5]{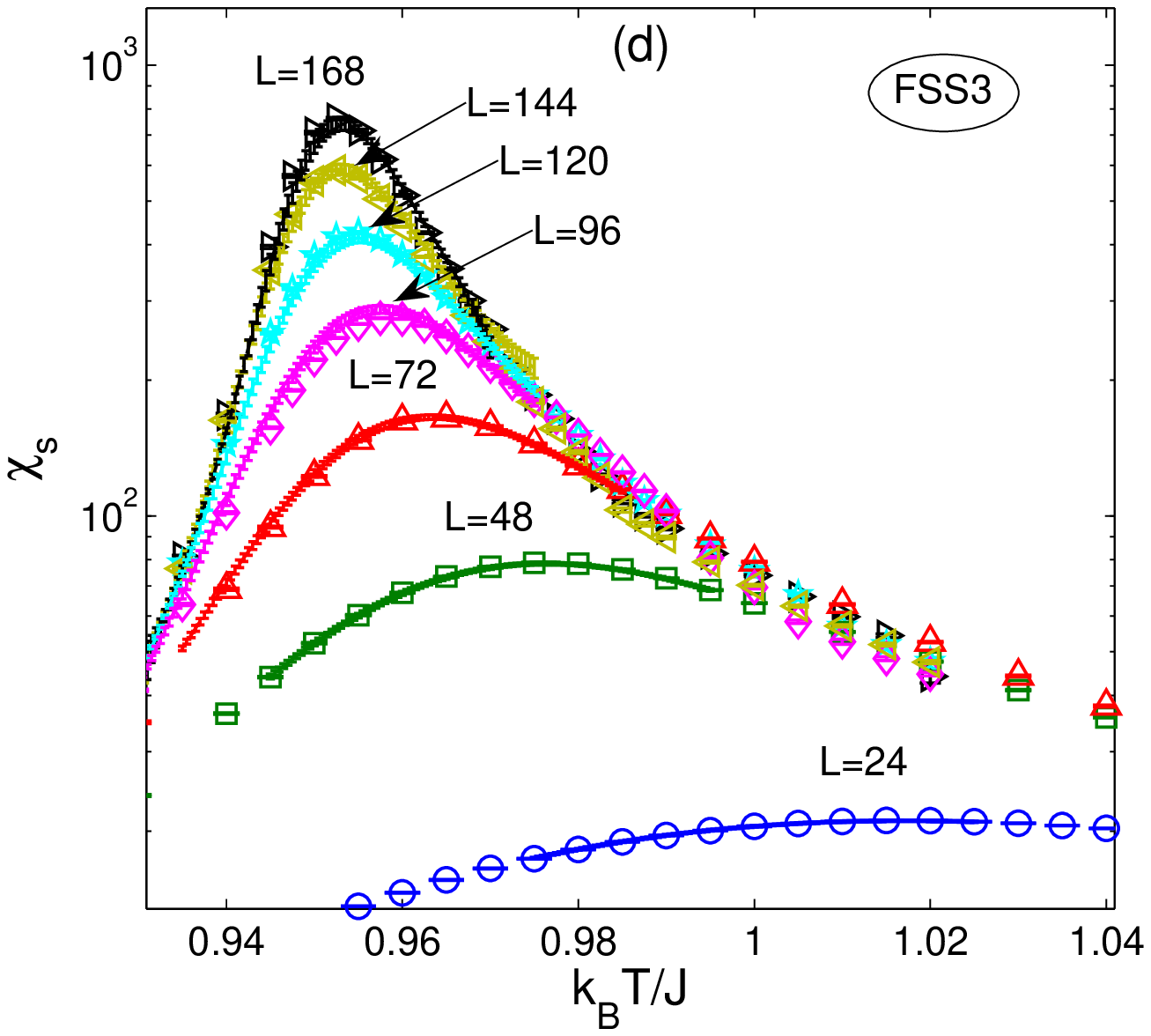}\label{fig:chis-T_L24-120_log_rew_pt}}
\caption{Temperature variations of the susceptibilities $\chi_x$, corresponding to the parameters $m_x$ ($x=\mathrm{A},\mathrm{B},s$), plotted for $L=24-120$. In (d) more detailed results in the vicinity of the critical point $T_{c1}$ are presented for the staggered susceptibility $\chi_s$ data obtained from the reweighing (dashed curves) and PT (symbols) techniques, for $L=24-168$.}\label{fig:T-x2}
\end{figure}

\begin{figure}[t]
\centering
		\subfigure{\includegraphics[clip,scale=0.5]{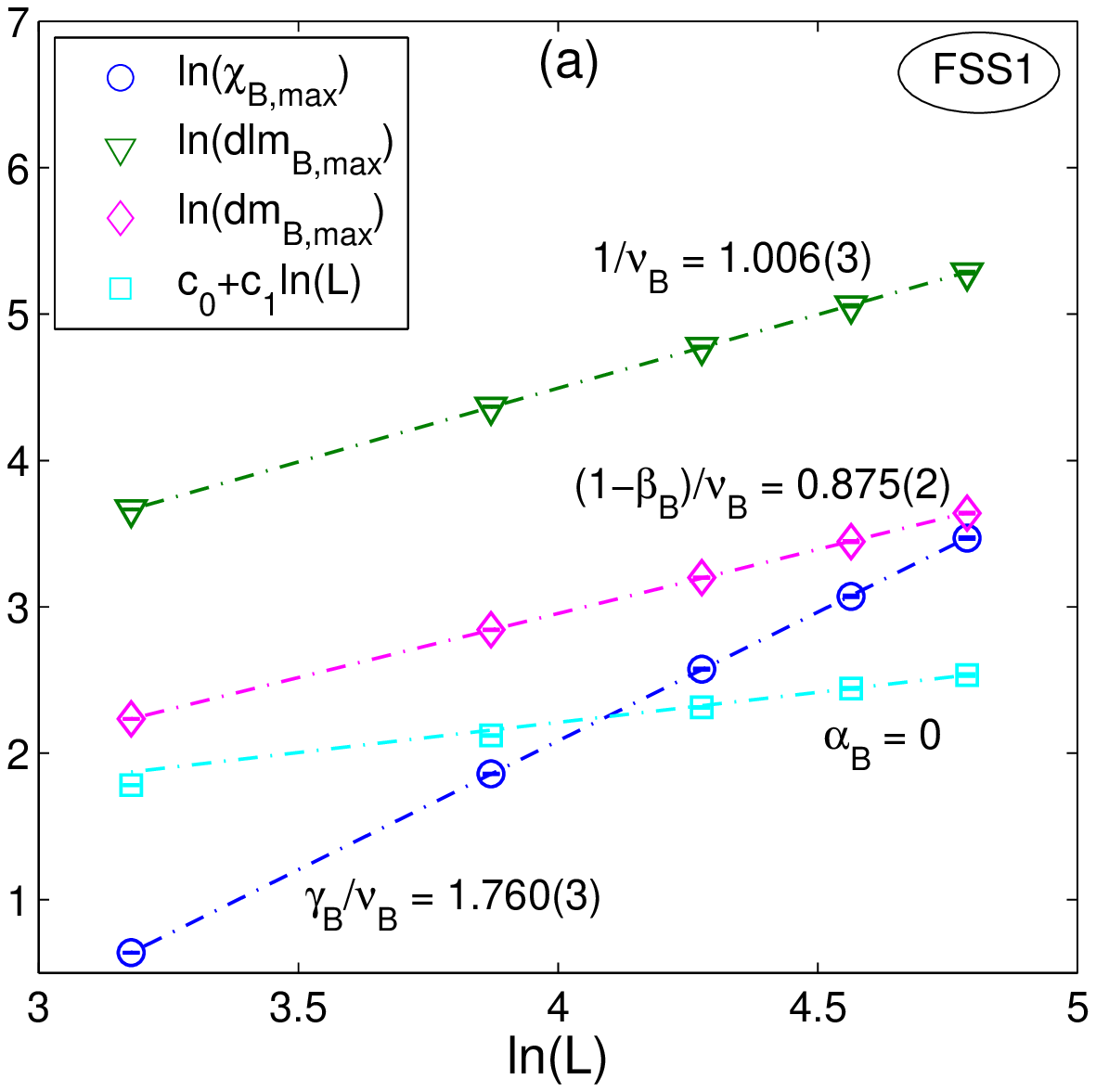}\label{fig:fss_T2_mB}}
		\subfigure{\includegraphics[clip,scale=0.5]{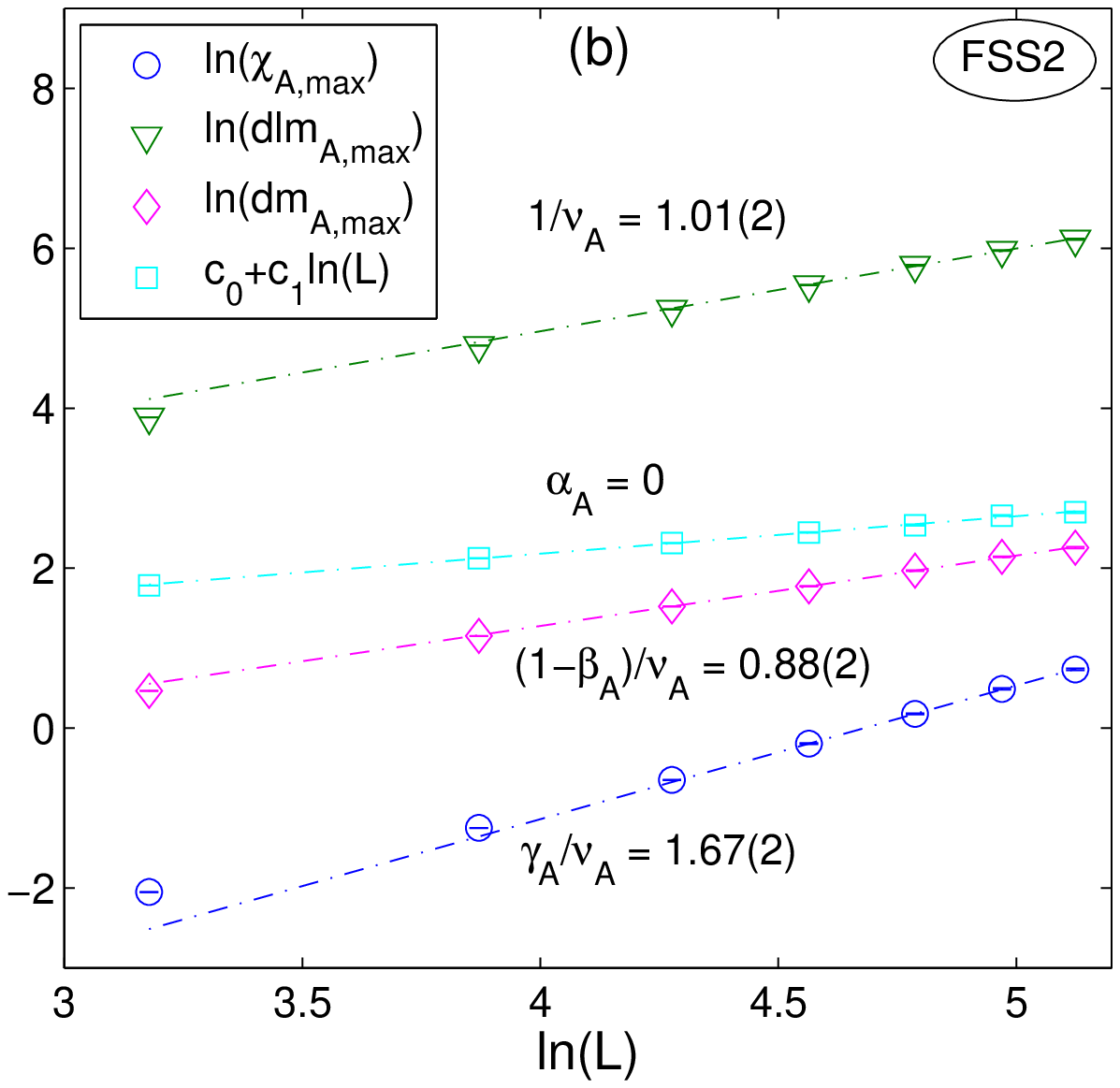}\label{fig:fss_T2_mA}}
		\subfigure{\includegraphics[clip,scale=0.5]{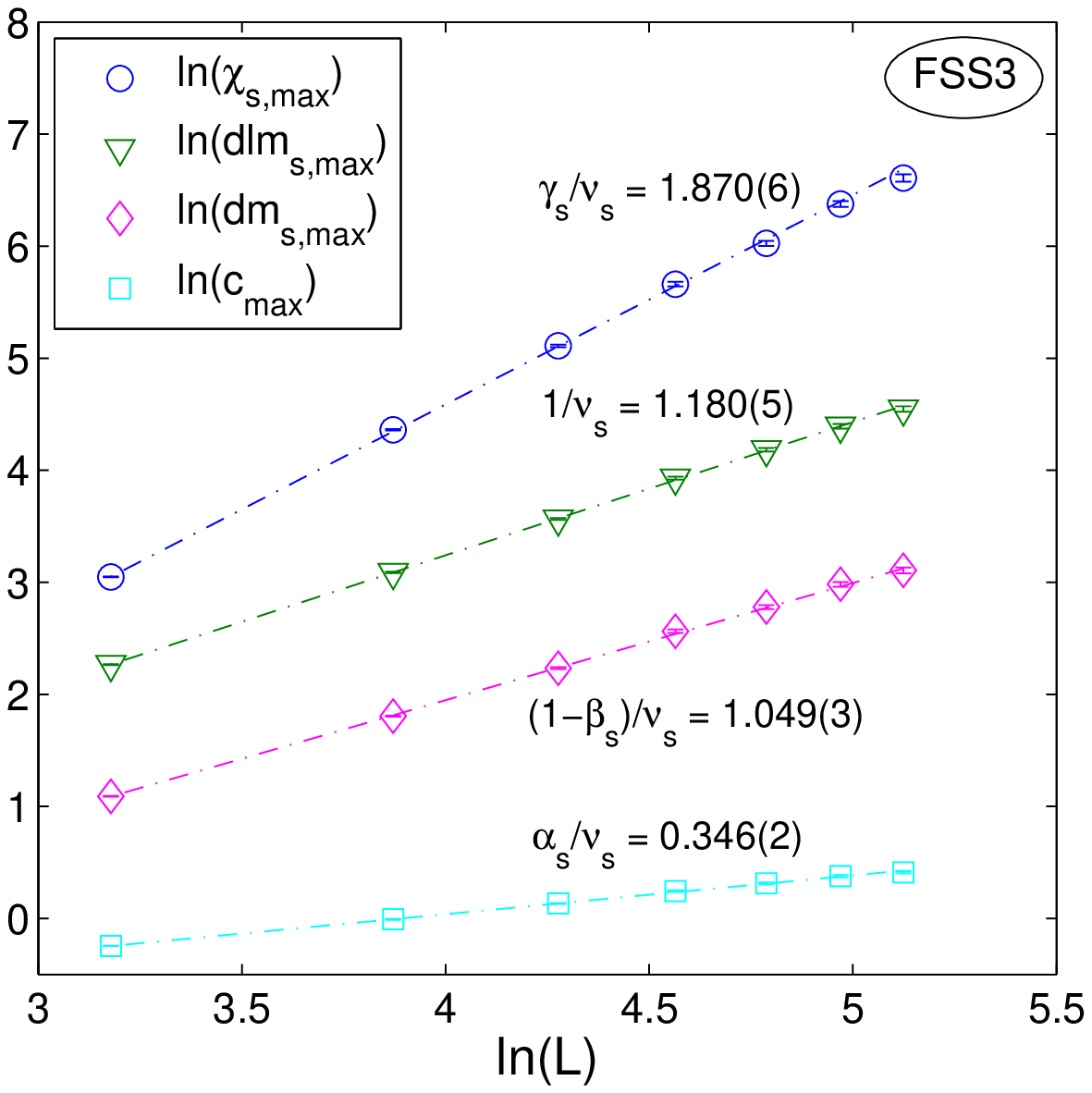}\label{fig:fss_T1_ms_new}}
\caption{Critical exponent ratios obtained in finite-size scaling analyses FSS1-3 from the scaling relations~(\ref{eq.scal_chi})-(\ref{eq.scal_c1}).}\label{fig:fss}
\end{figure}

\begin{figure}[t]
\centering
		\subfigure{\includegraphics[clip,scale=0.5]{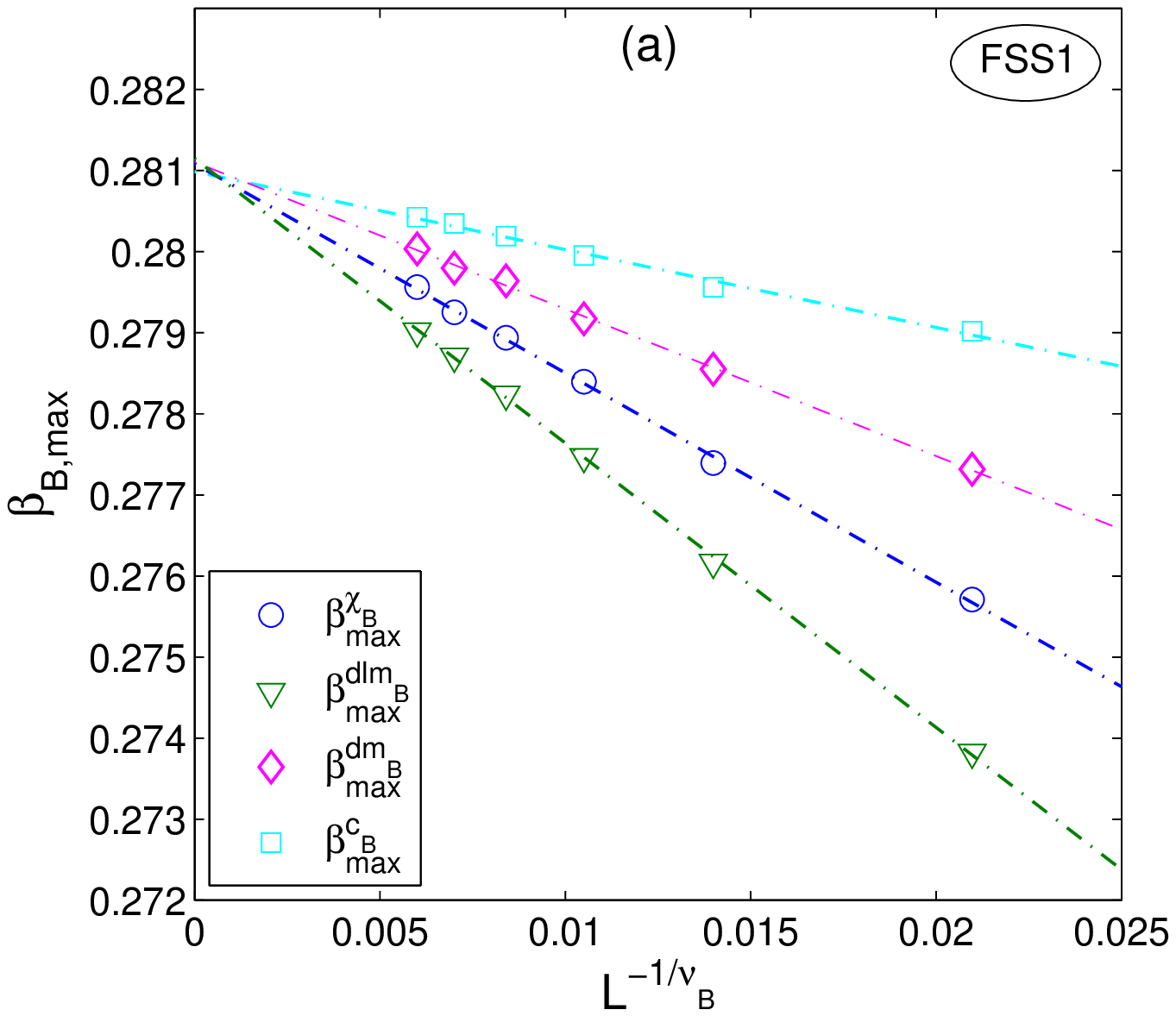}\label{fig:Tc_fss1}}
		\subfigure{\includegraphics[clip,scale=0.5]{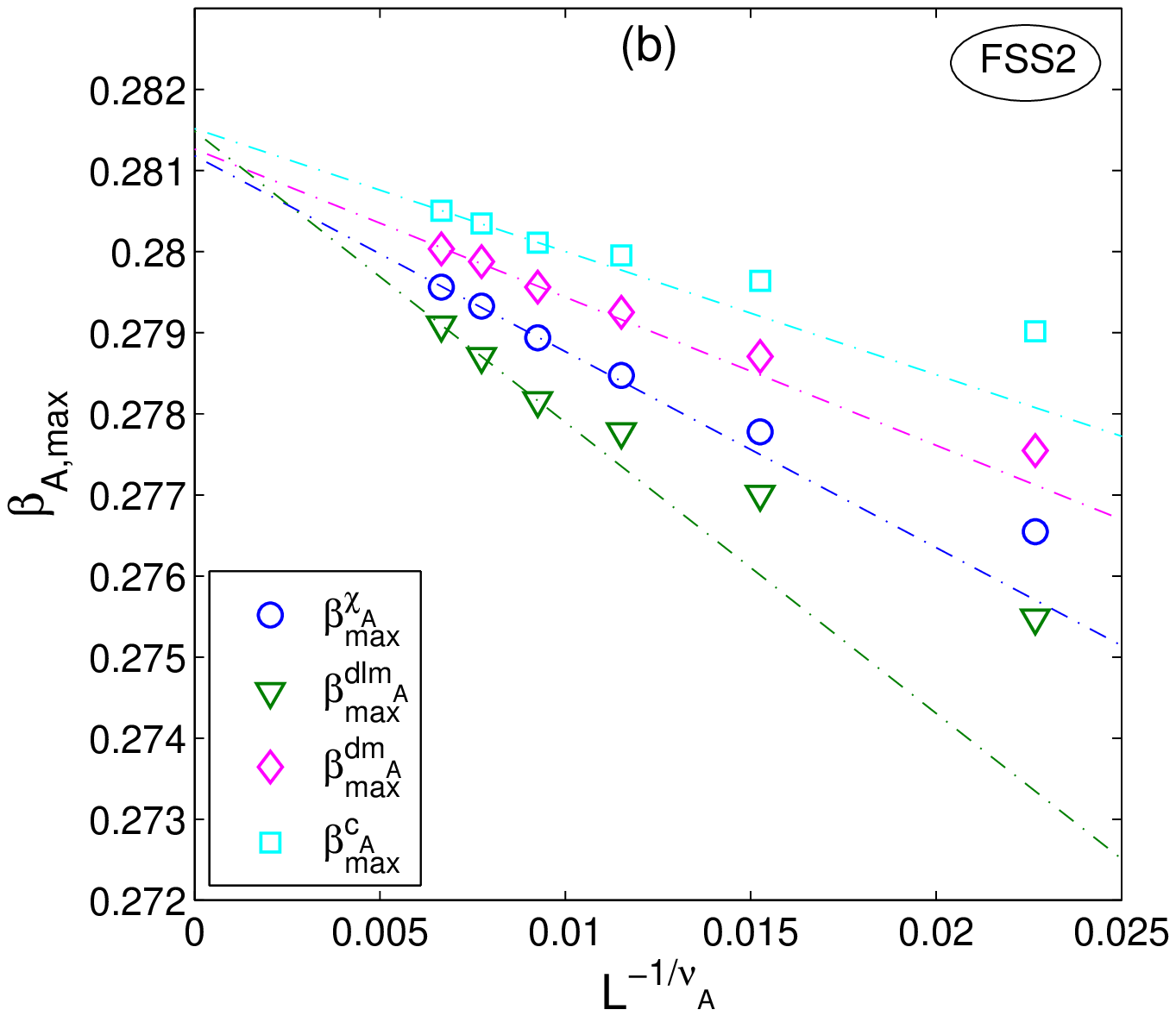}\label{fig:Tc_fss2}}
\caption{FSS fits of the inverse pseudo-transition temperatures $\beta_{x,max}$, according to the scaling relation~(\ref{eq.scal_Tc}), for the functions of $M_x$ and (a) $x=\mathrm{B}$ and (b) $x=\mathrm{A}$.}\label{fig:Tc_fss}
\end{figure}

\begin{figure}[t]
\centering
		\includegraphics[clip,scale=0.5]{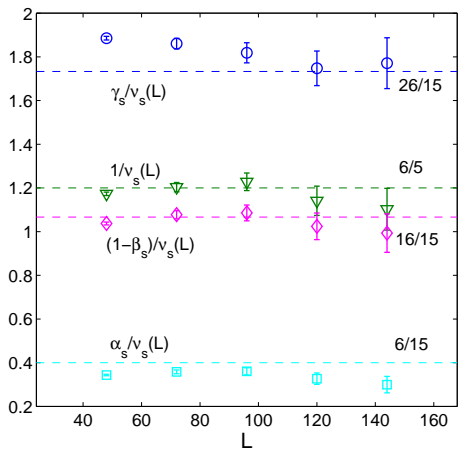}
\caption{Running exponents $\gamma_s/\nu_s(L),1/\nu_s(L),(1-\beta_s)/\nu_s(L),\alpha_s/\nu_s(L)$, for the low-temperature phase transition at $T_{c1}$. The dashed lines indicate the standard Potts values of the respective critical exponent ratios.}\label{fig:rol_fit}
\end{figure}

From the Hamiltonian~(\ref{Hamiltonian}) it is easy to verify that the ground-state arrangement of the system corresponds to the ferromagnetically ordered plane B with $m_{\mathrm{B}}=1$ and ferrimagnetically ordered plane A with two spins parallel and one antiparallel on each elementary triangle, which corresponds to $m_{\mathrm{A}}=1/3$ and $m_{s}=1$. Temperature dependencies of the calculated quantities, presented in Fig.~\ref{fig:T-x1}, provide the picture of the thermodynamic behavior of the system at finite temperatures as well as its dependence on the system size. In particular, in Fig.~\ref{fig:e-T} one can observe two anomalies in the internal energy dependencies associated with two phase transitions. Fig.~\ref{fig:m-T} shows that the high-temperature anomaly at some temperature $T_{c2}$ corresponds to the ferromagnetic (FM) long-range ordering (LRO) of the B plane, while the low-temperature one at $T_{c1} < T_{c2}$ reflects the onset of the ferrimagnetic (FRM1) LRO within the A plane, associated with the order parameter $m_s$. Nevertheless, one can notice that in the antiferromagnetic A plane there is a non-zero net magnetization $m_{\mathrm{A}}$ also within $T_{c1} < T < T_{c2}$ and is virtually independent on the lattice size. Such a ferrimagnetic state in the A plane is induced by the FM ordering in the B plane at $T_{c2}$ and we will refer to it as FRM2 phase.\\
\hspace*{5mm} The character of the respective phase transitions can be studied from the scaling of the corresponding response functions near criticality. The latter are shown in Figs.~\ref{fig:chiA-T_L24-120_log}-\ref{fig:chis-T_L24-120_log} for various $L$ on semi-log plots. At criticality one should expect linear dependence of their maxima with $\ln(L)$, according to the scaling relations~(\ref{eq.scal_chi})-(\ref{eq.scal_c1}), with the slopes (critical exponent ratios) providing information about the transition order and/or the universality class. The FSS analysis is performed using the above-defined quantities, corresponding to the parameters $m_x$ ($x=\mathrm{A},\mathrm{B},s$), as marked in Figs.~\ref{fig:chiA-T_L24-120_log}-\ref{fig:chis-T_L24-120_log} by FSS1-3. \\ 
\hspace*{5mm} However, the relatively large temperature step and the relatively small number of MCS used in standard MC simulations are not sufficient to locate the desired extremal values with high precision. Therefore, it is more convenient to resort to the reweighing techniques performed on much longer time series of the data obtained at the respective pseudo-transition points roughly located by the standard MC simulations. As already pointed out above, particularly simulations performed at lower temperatures, such as in the vicinity of the low-temperature transition temperature $T_{c1}$, involve some risk of getting stuck in local minima, which can be eliminated by the PT method. In Fig.~\ref{fig:chis-T_L24-120_log_rew_pt} we confront the results for the staggered susceptibility peaks obtained by the reweighing of the standard MC data and the PT methods. One can notice that even for the largest lattice sizes, which are the toughest to equilibrate, a fairly good agreement is achieved, thus boosting our confidence in reaching equilibrium solutions.\\
\hspace*{5mm} In Fig.~\ref{fig:fss} we present scaling results of the functions (\ref{eq.c})-(\ref{eq.D1}) at the high- and low-temperature transition temperatures $T_{c2}$ and $T_{c1}$. We note that $m_{\mathrm{B}}$ is the order parameter characterizing the high-temperature paramagnetic-ferromagnetic (P-FM) transition in the B plane and $m_s$ is the order parameter of the low-temperature transition to the ferrimagnetic FRM1 state in the A plane. However, as already indicated in Fig.~\ref{fig:m-T}, at both transitions there are abrupt changes also in the quantity $m_{\mathrm{A}}$ and the corresponding susceptibility (Fig.~\ref{fig:chiA-T_L24-120_log}) seems to diverge with the system size. \\ 
\hspace*{5mm} Indeed, all the log-log plots in Fig.~\ref{fig:fss} show power-law (or logarithmic in case of $c_{max}$) scaling, however, the respective phase transitions are governed by distinctly different critical exponent ratios. Those obtained for the P-FM transition in the B plane are consistent with the Ising universality class values: $\alpha_I=0$, $1/\nu_I=1$, $\gamma_I/\nu_I=1.75$ and $(1-\beta_I)/\nu_I=0.875$. Even the exponent ratios of the P-FRM2 transition in the A plane coincide within error bars with the standard Ising values, except for the susceptibility exponent $\gamma_\mathrm{A}$. One can notice that smaller lattice sizes do not comply with the power-law scaling and, therefore, we had to consider larger sizes up to $L=168$ and drop some smaller ones from the fitting. Nevertheless, the best fitted value $\gamma_\mathrm{A}/\nu_\mathrm{A}=1.67(2)$ is clearly smaller than the Ising value. The respective critical temperatures can be estimated from the scaling relation~(\ref{eq.scal_Tc}). As presented in Fig.~\ref{fig:Tc_fss}, the scaling is better behaved for the P-FM transition in the plane B (Fig.~\ref{fig:Tc_fss1}), while in the P-FRM2 transition the linear ansatz is only satisfied for $L \geq 120$ and the estimated values of the inverse critical temperature $\beta_c$ from different quantities are more scattered (Fig.~\ref{fig:Tc_fss2}). Nevertheless, the two transition temperatures cannot be distinguished within our limits of accuracy.\\ 
\hspace*{5mm} The low-temperature phase transition occurs in the A plane due to spin rearrangement from the partially ordered ferrimagnetic state FRM2 with $|m_{\mathrm{A}}| > 0$ to the ferrimagnetic state FRM1 with 2/3 of the spins aligned parallel and 1/3 antiparallel to the ferromagnetically ordered B-plane spins, resulting in the ground-state value of $|m_{\mathrm{A}}|=1/3$. Here, the role of the ferromagnetically ordered plane resembles that of an external magnetic field trying to align the spins in the antiferromagnetic A plane into its direction. The latter is believed to belong to the same universality class as the three-state ferromagnetic Potts model~\cite{alex75,kinz81} with the critical exponent ratios: $\alpha_P/\nu_P=6/15$, $1/\nu_P=6/5$, $\gamma_P/\nu_P=26/15=1.7\bar{3}$ and $(1-\beta_P)/\nu_P=16/15=1.0\bar{6}$. The present values obtained from the fits of the FSS3 analysis are rather close to the standard three-state Potts values, nevertheless, the values $\gamma_s/\nu_s=1.870(6)$ and $\alpha_s/\nu_s=0.346(2)$ apparently deviate beyond the error bars. However, even though the dependencies in Fig.~\ref{fig:fss_T1_ms_new} look linear a closer inspection reveals slight downward curvatures in the respective plots, indicating that the asymptotic regime may not have been reached. A more careful analysis involves evaluation of the running exponents $\gamma_s/\nu_s(L),1/\nu_s(L),(1-\beta_s)/\nu_s(L)$, and $\alpha_s/\nu_s(L)$, corresponding to local slopes estimated from three consecutive values of $L$. The results presented in Fig.~\ref{fig:rol_fit} indicate that for increasing $L$ also the ratio $\gamma_s/\nu_s(L)$ seems to converge to the Potts value, nevertheless, $\alpha_s/\nu_s(L)$ still remains below the standard value. We suspect that this deviation of $\alpha_s/\nu_s(L)$ might, at least partly, be caused by neglecting the nondivergent ``background'' term in Eq.~(\ref{eq.scal_c}). However, it is also possible that the present values indeed deviate from the standard Potts ones and show some variation with the strength of the coupling to the ferromagnetic layer. Similar behavior was also observed in the TLIA model the critical exponents of which displayed some dependence on the field value~\cite{kinz81,quier99}.   

\section{Conclusions}
Critical properties of an Ising bilayer spin system consisting of antiferromagnetic (AF) and ferromagnetic (FM) triangular planes, coupled by ferromagnetic exchange interaction, were studied by standard Monte Carlo and parallel tempering methods. At higher temperatures we identified in the FM plane a standard Ising universality class phase transition from a paramagnetic to a ferromagnetic state, which at the same critical temperature induces via the interlayer couplings a ferrimagnetic spin arrangement with non-vanishing magnetic moment in the adjacent AF plane. At lower temperatures, there is another phase transition in the AF plane to a different ferrimagnetic state with two sublattices aligned and one anti-aligned with the spins in the ferromagnetically ordered FM plane. This state resembles the ferrimagnetic arrangement of the TLIA model in moderate external magnetic fields~\cite{metcalf,schick,netz,zuko1}. The latter are believed to belong to the standard two-dimensional three-state ferromagnetic Potts universality class and the present results suggest that also the low-temperature phase transition in the present model belongs to same universality class. \\
\hspace*{5mm} Nevertheless, we think that it is likely that, similar to the TLIA model in a field, the critical exponents of the low-temperature phase transition in the present model can show some variation with the exchange interaction ratio. The effect of the exchange interaction anisotropy is of great interest also because it allows studying a dimensional cross-over phenomena in the system and may lead to qualitatively different critical behaviors. Such a study is currently underway.   

\section*{Acknowledgments}
This work was supported by the Scientific Grant Agency of Ministry of Education of Slovak Republic (Grant No. 1/0331/15).

\end{document}